\documentclass[12pt,useAMS,usenatbib,a4paper]{mn2e}

\usepackage{graphicx}
\usepackage{epsfig}
\usepackage{amssymb,amsmath}	
\usepackage{mathrsfs}
\usepackage[utf8x]{inputenc}
\usepackage[usenames]{color}
\usepackage{color}

\PrerenderUnicode{áéíóúñ}
\def\etal{et al. }

\title[BH Binaries Environment \& Properties]
{Supermassive Black Hole Binaries: Environment and Galaxy Host Properties of PTA and eLISA sources.}

\author[Mart\'inez Palafox \etal]{Eva Martínez Palafox$^{1}$\thanks{e-mail: evam@astro.unam.mx, eva.palafox@gmail.com} 
Octavio Valenzuela$^{1}$, Pedro Colín$^{2}$, Stefan Gottlöber$^3$
 \\
$^{1}$ Instituto de Astronomía, Universidad Nacional Aut\'onoma de M\'exico, A.P. 70-264
Ciudad Universitaria, D.F., M\'exico\\
$^2$ Centro de Radioastronomía y Astrofísica, Universidad Nacional Aut\'onoma de M\'exico, A.P. 72-3 (Xangari),
Morelia,\\ Michoacán 58089, México.\\
$^{3}$ Leibniz Institute for Astrophysics, An der Sternwarte 16, D-14482 Potsdam, Germany.\\
}

\begin{document}

\date{Received \today; in original form }

\pagerange{\pageref{firstpage}--\pageref{lastpage}} \pubyear{}

\maketitle

\label{firstpage}

\begin{abstract}
Supermassive black hole (BH) binaries would comprise the strongest sources of gravitational waves (GW) once they reach $\ll1$pc
separations, for both pulsar timing arrays (PTAs) and space based (SB) detectors. While BH binaries coalescences constitute a
natural outcome of the cosmological standard model and galaxy mergers, their dynamical evolution is still poorly understood and
therefore their abundances at different stages. We use a dynamical model for the decay of BH binaries coupled with a cosmological
simulation and semi-empirical approaches to the occupation of haloes by galaxies and BHs, in order to follow the evolution of the
properties distribution of galaxies hosting BH binaries candidates to decay due to GWs emission. Our models allow us to relax
simplifying hypothesis about the binaries occupation in galaxies and their mass, as well as redshift evolution. Following
previously proposed electromagnetic (EM) signatures of binaries in the subparsec regime, that include spectral features and
variability, we model possible distributions of such signatures and also set upper limits to their lifespan. We found a bimodal 
distribution of hosts properties, corresponding to BH binaries suitable to be detected by PTA and the ones detectable only from
space missions, as eLISA. Although it has been discussed that the peak of eLISA sources may happen at high z,  we show that there
must be a population of such sources in the nearby Universe that might show detectable EM signatures, representing an important
laboratory for multimessenger astrophysics. We found a weak dependence of galaxy host properties on the binaries occupation, that
can be traced back to the BH origin. The combination of the host correlations reported here with the expected EM signal, may be
helpful to verify the presence of nearby GW candidates, and to distinguish them from 'regular' intrinsic AGN variability. 

\end{abstract}


\section{Introduction}\label{Intro}
The current understanding of the cosmic large scale structure origin is included in the $\Lambda$CDM scenario (Planck Collaboration
\etal 2013). According to this cosmological model the self-gravitating systems at galactic scales have been formed following a
hierarchical process of mass accretion and mergers, and galaxies grew up by the collapse of baryons into the bottom of dark matter
haloes potential wells (White \& Rees 1978; Rees  \& Ostriker 1977; Silk \& Mamon 2012 and references therein). Therefore it is
possible to find a relationship between dark matter haloes mass accretion/merging histories and the interactions between galaxies. The
consistency between the halo mass function and the galaxy luminosity function predicts a relationship between the halo mass and the
observed galaxy stellar mass, and it is at the core of the halo occupation distribution (HOD) techniques, widely used to test the
consistency between galaxies population and $\Lambda$CDM cosmology.

On the other hand, stellar kinematic studies of nearby galaxies (Kormendy \& Richstone 1995; Richstone et al. 1998; Ferrarese \& Ford
2005) suggest that a supermassive black hole reside at their centres. Currently, our Milky Way (MW) is the most striking example (Ghez
et al. 2008; Gillessen et al. 2009). There is also a variety of investigations confirming that the mass of the BH is highly correlated
to the mass and velocity dispersion of the host galaxy bulge  (Magorrian et al. 1998; Ferrarese \& Merritt 2000; Gebhardt et al. 2000;
Marconi \& Hunt 2003; Häring \& Rix 2004; Graham et al. 2011; McConnell \& Ma 2013). Additionally, still controversial correlations
between total luminosity or mass and BH mass (Ferrarese 2002; Baldana \etal 2009; Läsker \etal 2013) suggest that BH and their host
galaxies evolve in a symbiotic way (Cattaneo \etal 2009; Kormendy \& Ho 2013). A natural consequence is that from galaxy mergers, a
large number of BH binaries must form along the cosmic history (Begelman, Blandford \& Rees 1980; Volonteri, Haardt \& Madau 2003).
However, few spatially resolved subkpc binaries are known so far (Rodriguez et al. 2006; Fabbiano et al. 2011; Woo et al. 2014), and
the few observations of sub parsec candidates reported (Tsalmanza \etal 2011; Eracleous 2012; Liu \etal 2014) could also be explained
by a number of scenarios other than the presence of a binary (Dotti \etal 2012). 

The search for and characterization of the corresponding binary BH population are of great importance to understand both galaxy
formation and gravitational wave (GW) physics. The dynamical evolution of the binary system within the merged galaxy, and the
interaction with the host both dynamically and  possibly via feedback during baryonic accretion onto one or both BHs, encode crucial
information about the assembly of the bulge and the central BH. If the binary eventually coalesces, it will produce a gravitational
siren that could be detected with future low frequency GW experiments (e.g., Colpi \& Dotti 2011). Detecting BH binaries is then
crucial in understanding the coevolution of galaxies and BHs, and for testing fundamental physics. The BH binaries may produce a wide
range of EM signals in their path to coalesce: Emission associated to a circumbinary disk (CBD) (Milosavljević \etal 2005; Tanaka \&
Menou 2010; Tanaka \etal 2012; Sesana \etal 2012; McKernan \etal 2013; Roedig \etal 2014), variability due to the orbital motion (Gower
\etal 1982; Sillanpaa \etal 1988; Kaastra \& Roos 1992; Britzen \etal 2010; Kudryavtseva \etal 2011; Godfrey \etal 2012), multiple
accreting BHs spatially resolved (Komossa \etal 2003; Rodriguez \etal 2006; Fabbiano \etal 2011; Koss \etal 2011; Paggi \etal 2013),
enhanced tidal disruption events (Ivanov \etal 2005; Chen \etal 2009;  Wegg \& Nate Bode 2011) and BHs wandering as a result of
three-body interactions (Governato \etal 1994; Iwasawa \etal 2006) or post-merger recoiling BHs (Komossa \& Merritt 20008; Merritt
\etal 2009; O’Leary \& Loeb 2009). 

The frequency window of GW from BH binaries corresponds to the low frequencies in the range ($10^{-2} - 10^{-9}$)Hz covered by two
projects: eLISA and PTAs. The \textit{Evolved Laser Interferometer Space Antenna\footnote{https://www.elisascience.org/}} (eLISA) 
will be the first dedicated space based gravitational wave detector. It will measure gravitational waves at low frequencies, from about
$0.1$ mHz to $1$ Hz (Amaro-Seoane \etal 2013). The potential \textit{eLISA} sources include BH close binaries in the process of
coalescence, with total mass in the range $M_{\rm BH}=M_{\bullet,1}+M_{\bullet,2}  \sim 10^4 - 10^7 \text{M}_{\odot}$ (where
$M_{\bullet,1}$ and $M_{\bullet,2}$ are the primary and secondary BH masses respectively). The technique to detect GW of ultra-low
frequencies ($\sim10^{-9}$ to $10^{-8}$ Hz) is called Pulsar Timing Arrays (PTAs), consisting of sets of millisecond pulsar whose joint
pulse arrival time produce correlations that have encoded information of gravitational waves passing the Earth and/or the pulsar
(Foster \& Backer, 1990). PTAs will detect GW from the coalescence of BH binaries with masses $M_{\rm BH} \gtrsim10^8 M_\odot$ and mass
ratio $10^{-3} \lesssim q = M_{\bullet,2}/M_{\bullet,1}\lesssim 1$ (Sesana \etal 2012).

Before we reach the direct GW detection era, the possibility of detecting EM counterparts revealing binary coalescence events
unambiguously has valuable potential, either constraining the dynamics of the coalescence process (e.g., Sesana 2007; Trias \& Sintes
2008; Amaro-Seoane et al. 2012), or the origin and evolution of BHs (e.g., Volonteri 2010, Kulkarni \& Loeb 2012), and the environments
where these events take place. Even though the richness offered by the EM signatures produced in binaries the aim is to reach the
multimessenger astrophysics era of trans-spectral studies (GW+EM) to narrow the degeneracies of a possible GW detection (e.g., Bloom 
\etal 2009). Recent studies have aimed to narrow this multimessenger search, Sesana \etal (2012)  used a semi-analytical galaxy
formation model and a massive BH halo population algorithm in order to predict the frequency of different EM signals related with PTA
sources. Later Rosado \& Sesana (2013) study the environment of BH binaries suitable to be detected by PTAs, finding a bias to high
density environments. Lastly Simon \etal (2014) assuming that each galaxy has a binary BH of equal mass, evaluated probable locations
of PTA sources, suggesting nearby clusters.

In this work we intend to relax some of the assumptions adopted in previous studies, we build a more realistic binary population and we
do not assume that all binaries in the Local Universe are in the same coalescence stage, however we do not try to model the
detectability. By focusing the modelling of binaries since the last major merger (LMM) of their host galaxies and following their
dynamical evolution, we expect to understand better those objects that are more likely to be reliably considered galaxy candidates to
host binaries. In order to do so, we populate under different strategies the dark matter haloes in a cosmological simulation during the
last major merger event and track down the binary evolution, including the interaction with stars and a gaseous circumbinary disk. We
use halo occupation distribution (HOD) techniques and semiempirical halo/galaxies scaling relationships (Firmani \& Ávila-Reese 2010)
in order to assign a BH and galaxy properties to each halo, satisfying scale relationships such as luminosity function, in order to
asses the binary BH galaxy host different properties with more accuracy. The assignment was performed at different epochs in order to
track down binary galaxy host properties  at different redshifts. We estimate the lifespan and variability of electromagnetic signals.

This paper is organized as follows. 
In Section 2 we describe the parameters and analysis adopted for our cosmological simulation. 
In section 3 we describe the prescriptions implemented in order to populate dark matter haloes with stellar disks and BHs.
In section 4 we describe the semi-analytical model implemented to follow the dynamical evolution of BH pairs from cosmological scales
down to subparsec scales. 
In section 5 we discuss the properties found for galaxies hosting binaries with separations $~10^{-3}$pc, candidates to decay due to GW
emission.
In section 6 we classify binaries in the previous section into PTA and SB sources, then we present the bimodality found in host galaxy
properties, as total mass, stellar mass, and environment.
In section 7  we present the assumptions made in order to assign a possible EM signature, then we discuss the abundance and
characteristics of the EM features that might be found in the binary populations resulting from our models.
Finally, we summarize the main conclusions of this paper in Section 8. 
\section{Cosmological simulation.}
 \begin{figure*}
  \centering
  \includegraphics[width=\linewidth]{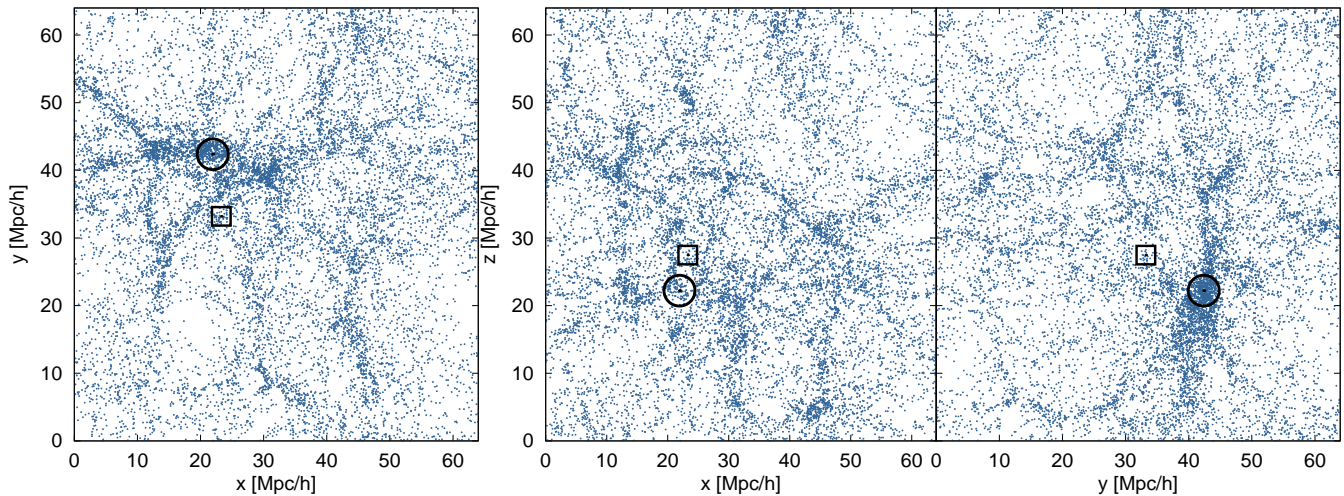}
  \caption{Halo distribution in the CLUES simulation. Only haloes more massive than $M_h  \geq 5\times 10^{10} \text{M}_{\odot} h^{-1}$
are shown as blue dots. The black circle corresponds to the projected position of the halo identified with the Virgo cluster. The black
square corresponds to the projected position of the halo dubbed as \textit{MW candidate}.}
  \label{fig:mapatot}
 \end{figure*}

In this work we use a cosmological N-body dark matter simulation which is part of the Constrained Local UniversE Simulations (CLUES)
project,\footnote{http://www.clues-project.org/} whose aim is to perform cosmological simulations that reproduce the main features that
characterize the local large scale structure in the Universe (Gottlöeber \etal 2010). Figure \ref{fig:mapatot} shows the distribution
in the CLUES simulation of haloes more massive than $M_h  \geq 5\times 10^{10} \text{M}_{\odot} h^{-1}$, shown as blue dots. It also
shows the projected positions of the halo identified with the Virgo cluster and the halo dubbed as \textit{MW candidate} as black 
circle and black square, respectively. The Hoffman \& Ribak (1991) algorithm was used to generate the initial conditions as constrained
realizations of Gaussian random fields. Constrained simulations represent a powerful tool to understand the history of the best  known
part of the Universe. By using them it is possible to construct a constrained sample of possible merging galaxies corresponding to any
epoch of the local Universe. Also, constrained simulations allow us to explore the assembly history of MW and M31-like galaxies
(Forero-Romero \etal 2011), considered as candidates for such systems due not only to theirs masses (as usually assumed in other dark
matter simulations) but also because their environments and MAHs have been shaped along the cosmic history by the large scale structure
in which the Local Universe actually dwells. Advantages we will use to explore the origin and evolution of the BHs in galaxies in the
Local Group in an upcoming paper. We summarize the simulation main properties here for clarity.

The simulation has $1024^3$ particles with mass $m_p= 1.63\times10^{7}h^{-1} \text{M}_{\odot}$ in a box with side length of $L_{box} =
64 h^{-1}$Mpc. It was run using the PMTree-SPH MPI code \verb|Gadget|2 (Springel 2005). The simulation has standard $\Lambda$CDM
initial conditions, assuming a WMAP3 cosmology (Spergel et. al. 2007), i.e. $\Omega_m=0.24$, $\Omega_{b}=0.042$,
$\Omega_{\Lambda}=0.76$, a normalization of $\sigma_8=0.75$ and a spectral index of primordial density perturbations  $n=0.95$.

As a frame, our model uses mass aggregation histories (MAHs) from the simulation to identify the last major merger redshift (zLMM) and
to estimate the \textit{cosmic mass accretion rate} which corresponds to the mass accreted going to the very centre of the galaxy. The
impact of cosmic variance on the CLUES simulation's halo population has been studied before (Forero-Romero et. al. 2011). They conclude
that constrained simulation MAHs are essentially unbiased when compared to unconstrained simulations with higher resolution within
larger volumes.

\subsection{Halo identification and mass aggregation histories construction.}
To identify haloes we use a friends-of-friends (FOF) algorithm with linking length of $b=0.17$ times the mean inter-particle
separation. We have identified the haloes for 133 snapshots more or less equally spaced over the 13.4 Gyrs between redshifts $0 < z
\lesssim 13.9$. All objects with 20 or more particles are kept in the halo catalogue and considered in the merger tree construction.
This corresponds to a minimum halo mass of $M_{min}= 3.26\times 10^{8} h^{-1}\mathrm{M}_{\odot}$. Within this simulation a Milky Way
dark matter halo of mass $10^{12} h^{-1}\mathrm{M}_{\odot}$ is resolved with $\sim6\times10^4$ particles.

The MAHs do not include any information of the substructure in each halo. MAHs are constructed by comparing the particles in FOF groups
in two consecutive snapshots. Starting at $z=0$ for every halo $\rm H_{f}$ in the catalogue, we find all the FOF groups in the previous
snapshots that share at least 13 particles with the halo $\rm H_{f}$ at $z=0$, and label them as potential progenitors. Then, for each
tentative progenitor, we find all the descendants sharing at least 13 particles. Since the smallest group has 20 particles, at least
$2/3$ of the particles must be identified in provisional progenitors or descendants. Only the tentative progenitors whose main
descendant is $\rm H_{f}$, are labelled as confirmed progenitors at that snapshot. We iterate this procedure for each confirmed
progenitor, until reaching the last available snapshot at high redshift. By construction, each halo in the MAH can have only one
descendant, but many progenitors.

The time at which haloes mergers occur corresponds to the moment when their radii overlap for the first time. Major mergers (MM) are
defined in terms of an increase of at least $20\%$ in the total mass of a descendant halo. The LMM time is identified in the MAHs for
each halo with mass $M_{\rm H} \geq 1.64 \times 10^{10} M_\odot h^{-1}$ (at $z=0$). We verify the reliability of the merger event
inspecting the  neighbour snapshots. The zLMM provides a starting point for our model, and we consider it the first stage in the
coalescence process.

\section{Populating haloes with galaxies and black holes} 

Our N-body simulation includes only dark matter, if we intend to constraint the properties of binary BH host galaxies we require a
strategy to populate dark matter halos with galaxies, below we describe our galaxies/black holes assignment scheme.

\subsection{Seeding Galaxies}
The galaxy/dark halo connection has received an important deal of attention in recent years. Such connection have been commonly studied
using \textit{ab initio} strategies: N-body simulations coupled with semi-analytic galaxy formation models or self consistent
cosmological hydrodynamic plus N-body simulations. An alternative strategy with an empirical spirit is to assume a direct match between
dark halos and observed galaxy properties, in particular stellar mass. Although this approach does not determine the physical processes
behind the correspondence, it can be extended even to trace the galaxies halo/stellar mass evolution assuring a consistency with a set
of observations. We use the latter option, a semi-empirical approach to define a galaxy population consistent with current
observations, including: luminosity function and downsizing (Firmani \& Ávila-Reese 2010). The implementation assumes that we can
transform the dark halo MAHs into average stellar mass growth histories, $M_s(M_{\rm H}(z),z)$, following the semi-empirical relation
given by Firmani \& Ávila-Reese (2010), 
\begin{equation}
 \begin{split}
 \mathrm{ log}(M_{\rm H}(M_s)) &= \mathrm{ log}(M_1) +  \beta \mathrm{ log} \left( \frac{M_s}{M_{s,0}} \right)  \\
& + \frac{\left( \frac{Ms}{M_{s,0}} \right)^{\delta}
}{1+\left( \frac{M_{s,0}}{M_s} \right)^{\gamma} } - \frac{1}{2}, \label{vladi}
 \end{split}
\end{equation}
where the dependence on $z$ is introduce in the parameters of equation \ref{vladi} as
\begin{equation*}
 \begin{array}{l c l}
 \mathrm{ log}(M_1(a))     &=& M_{1,0} + M_{1,a}(a-1),                \\
 \mathrm{ log}(M_{s,0}(a)) &=& M_{s,0,0} + M_{s,0,a}(a-1) + \chi(z),  \\ 
\label{paravladi}
  \delta(a)                &=& \delta_0  + \delta_a (a-1),            \\
 \chi(z)                   &=& -0.181z(1-0.378z(1-0.085z)),            \\
 \end{array}
\end{equation*}
whose values were chosen according to table \ref{vladt}. $M_{\rm H}$ is the total mass of the host galaxy and $a=1/(1+z)$ is the scale
factor.
\begin{table}
\centering
 \begin{tabular}{l c r}
\hline
Parameter & &value \\
 \hline
  $ M_{s,0,0}$    & & $10.70 $        \\
  $ M_{s,0,a}$    & & $-0.80 $        \\
  $ M_{1,0}$      & & $ 12.35$        \\
  $ M_{1,a}$      & & $ -0.80$        \\
  $ \beta$        & & $ 0.44 $        \\
  $ \delta_0$     & & $ 0.48 $        \\             
  $ \delta_{a}$   & & $-0.15 $        \\
  $ \gamma$       & & $ 1.56 $        \\
\hline

 \end{tabular}
\caption { \small  Best parameters for the $M_{\rm H}(M_s, 0<z<4)$, total mass-stellar mass relation (Firmani \& Ávila-Reese 2010).}
\label{vladt}
\end{table}
Now, provided the total mass of the galaxy, we have to associate a mass density distribution. This is a complicated and unfinished task
in galaxy formation theory. Providing a prescription to choose between spiral and elliptical morphology and their structural parameters
is an active field in semi-analytical models and hydrodynamic galaxy formation experiments. As a working simplification hypothesis we
associated to each remnant halo a scale radius for late type galaxies as given by Shen \etal (2003) 
\begin{equation}
 R(kpc) = 0.1 \left(\frac{M_{\rm H}}{M_{\odot}}\right)  ^{0.14}
\left(1+\frac{M_{\rm H}}{3.98\times10^{10}M_{\odot}}\right)^{0.25}.
\end{equation}
This assumption allowed us to model galaxies' potentials for each remnant halo using a Miyamoto-Nagai disk (1975) and a NFW halo (1996)
in order to investigate the binary BH decay due to angular momentum exchange with stars at kpc scales. The assumption is not accurate,
however as we can see in figure 2, last major merger redshift for our binaries is  clustered around 1, the typical gas richness at that
epoch favours disk reformation in the remnant galaxy (Robertson et al. 2006). Assuming a disk/spheroid component is more realistic,
however assigning relative parameters is highly uncertain from theoretical grounds (e.g. Ávila-Reese et al. 2014). The presence of a
bulge might shorten the time scale determined by the angular momentum exchange with stars and some binaries stalled at this stage might
reach the phase of decay due to GW emission. In this paper our main focus are the galaxy hosts properties of BH binaries at the GW
regime and not the exact abundance, therefore we consider that the effect is not critical for our results.

\subsection{Seeding black holes}

We used two different approaches in order to provide a BH mass appropriate to the characteristics that each progenitor had before the
merger, and to investigate the BH-halo occupation effects in the final properties of the binaries populations. First, we used the
empirical relation between black hole mass and the total gravitational mass of the host galaxy suggested by Ferrarese (2002),
\begin{equation}
 M_{\bullet} = 0.67 \left( \frac{M_{\rm H,i}}{10^{12} M_{\odot}} \right)^{1.82}, 
\end{equation} \label{Fbhmass}
where $M_{\bullet}$ is the mass of the BH, in $10^8 M_{\odot}$ units and $M_{\rm H,i}$ the total mass of the progenitor galaxy at hand.
Models with a BH seeded following the Ferrarese (2002) relationship will be called F. 


Second, an approach based on HOD calibrated by hydrodynamic cosmological simulations that follows BH growth (Di Matteo \etal 2008).
Despite the uncertainties in AGN physics and accretion mechanisms, these simulations reproduce both the observed $M_{\rm BH}-\sigma$ 
relation and total BH mass density $\rho_{\rm BH}$ (Di Matteo \etal 2008), as well as the quasar luminosity function (Degraf \etal
2010). In our model, the BH population seeded in dark matter haloes following this approach inherits such properties. The conditional
BH mass function (CMF) gives the mean number of BHs per logarithmic BH mass bin found in haloes of mass $M_{\rm H}$ and is fitted by a
Gaussian distribution (Degraf \etal 2011), in $log_{10}(M_{\bullet})$.
\begin{equation}
 \frac{dN_{\rm BH}}{d logM_{\bullet}} = \frac{1}{\sqrt{2\pi \sigma^2}} e^{-\frac{(log(M_{\bullet})-\mu)^2}{2\sigma ^2}}
\end{equation}
where the parameters $\sigma(M_{\rm H})$ and $\mu(M_{\rm H})$ corresponds to their best fit (Degraf \etal 2011). Models with a BH
seeded following the CMF in Degraf \etal (2011) will be called O hereafter.

\begin{figure*}
 \centering
 \includegraphics[width=\linewidth]{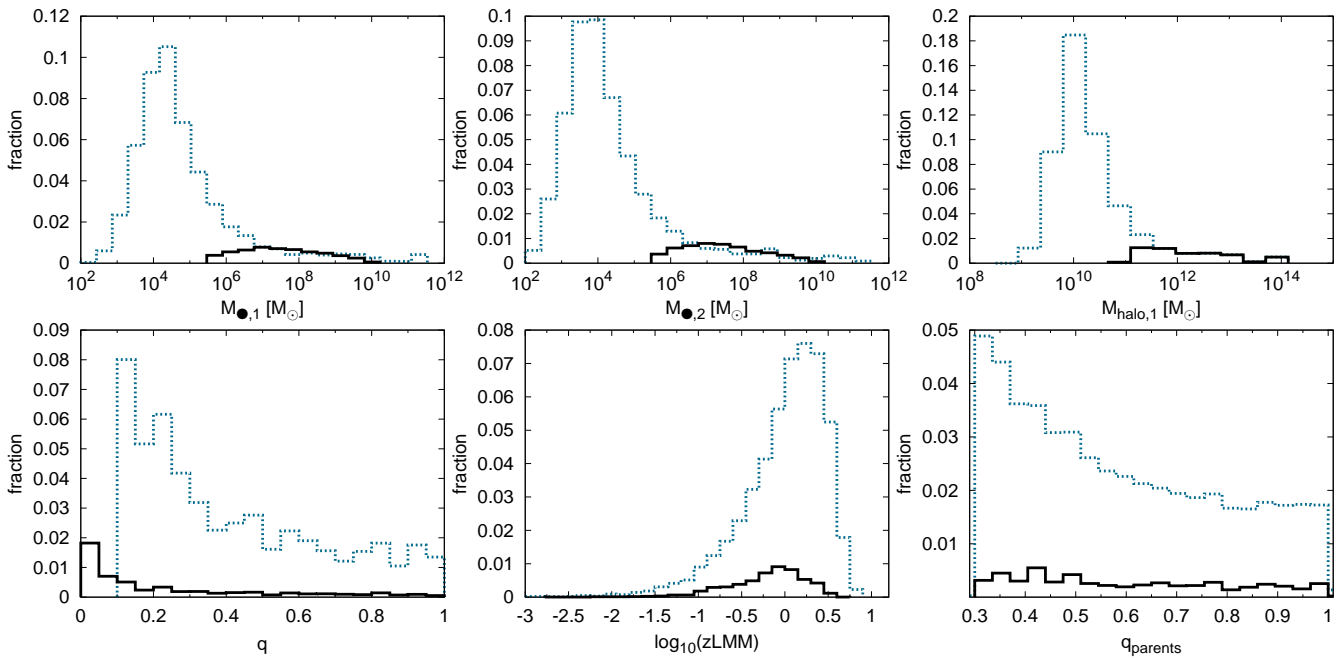}
 \caption{Primary and secondary BH mass, primary halo mass, BH mass ratio, last major merger redshift and progenitors mass ratio
initial distributions used in the 10 parameter sets explored (see table \ref{parametros}). Blue dotted lines correspond to F seeding
prescription (empirical $M_{\rm H} - M_{\rm BH}$ relation), which consists of 24815 galaxy pairs, black solid lines correspond to
the O seeding prescription (occupation model+simulation based) consisting of 2680 galaxy pairs. The difference in the number of
pairs is due to the fact that O prescription only works for haloes with masses between $10^{11}\: -\: 10^{14}\:M_{\odot}$ as can
be seen in the third top panel.}
 \label{fig:IC}
\end{figure*}

Figure \ref{fig:IC} shows the  primary and secondary BH mass, primary (or main progenitor) halo mass, BH mass ratio, last major merger
redshift and progenitors mass ratio initial distributions used in the 10 parameter sets explored (see table \ref{parametros}). Blue
dotted lines correspond to the $M_{\rm H}-M_{\rm BH}$ empirical BH seeding prescription F and consists of 24815 galaxy pairs, black
solid lines correspond to the BH seeding prescription based on occupation models O and consists of 2680 galaxy pairs. The difference in
the number of pairs is due to the fact that the O seeding prescription is defined only  for haloes with masses  between $10^{11}\: -\:
10^{14}\:M_{\odot}$ as can be seen in the third top panel. Also, given the restriction imposed to the halo progenitor's mass ratio of
$q_{p} \geq 0.3$, the F seeding prescription constrains the minimum possible binary BH mass ratio $q$ to be $0.11$, as seen in figure
\ref{fig:IC} first bottom panel, whilst O prescription allows all values of $q$ between $0$ and $1$ given its broader scatter. It is
important to mention that not all populated halos will reach the decay regime  dominated by GW emission.

Given the differences between seeding approaches, they might be considered as the consequence of different BH formation scenarios.
Then, by using two approaches we are able to investigate their effect in the distribution of galaxy hosts properties at the end of the
binary evolution. If there is a significant effect, it may set an avenue to constrain BH formation scenarios that depends on the binary
population.

\section{Dynamical Decay Model Description}
In our dynamical coalescence model, supermassive black hole binaries are considered to evolve via four stages, identified by the
separation between BHs, from large to small separations, and the main interactions with the surrounding material. The four stages
considered are (1) the last major merger epoch obtained from a constrained cosmological simulation, as mentioned in section 2; (2)
time scale for decay of merging galaxies' baryonic components within a remnant halo, (3) dynamical friction of the secondary BH
$M_{\bullet,2}$ to reach the centre of the newly formed galaxy, and (4) binary-gaseous disk interaction until the GW driven decay
regime is reached. Here we introduce the dominant dynamical mechanisms that we have considered drive the binaries evolution.

\subsection{Merger time scale for galaxies within the remnant halo}
The halo merger time, obtained from dark matter only simulations of large cosmological boxes, is frequently underestimated. Typically,
it is defined when the satellite halo has crossed the host's virial radius. This definition only includes the beginning of the merger
and it does not take into account the coalescence time needed to have a new remnant halo. Additionally, because of the large simulated
volume, the mass and spatial resolution are not always enough to fully capture all the density modes related to dynamical friction.
These inaccuracies have stimulated the inclusion of some corrections, for instance: those based on the analytical description of
dynamical friction given by Chandrasekhar or modified versions due to finite halo size and density gradients, or angular momentum (see
Zavala \etal 2012 and references therein). Given this uncertainties, we have chosen all haloes with known progenitors at their LMM and
kept only those with progenitors' mass ratios between $0.3 \leq q_{p} =M_{\rm H,2}/M_{\rm H,1} \leq 1$. Since Boylan-Kolchin, Ma \&
Quataert (2008) in their N-body experiments showed that, for such mass ratios, the coalescence time scale for an extended satellite to
sink from the host halo virial radius down to the centre, is similar to the host halo dynamical time scale. This  time scale is much
longer for more dissimilar mass ratios. Thus, the time scale for the second stage in our model for the coalescence process is given by
the dynamical time scale at the virial radius of the main progenitor,
\begin{equation}
 \tau_{dyn} \equiv \frac{r_{vir}}{V_c(r_{vir})} = \left( \frac{r_{vir}^3}{GM_{\rm H,1}} \right)^{1/2},
\end{equation}
where $r_{vir}$ is the virial radius, $V_c(r_{vir})$ is the circular velocity at $r_{vir}$, and $M_{\rm H,1}$ is the halo mass before
the merger. 

\subsection{Binary interaction with stars}
In the third stage, the black holes sink into the potential well depth given by the remnant galaxy due to dynamical friction with stars
and dark matter.

At this point we will start studying the dynamical evolution of the BHs from the host and the satellite haloes. The deceleration
experimented by the black hole embedded in stars and dark matter can be provided by the Chandrasekhar dynamical friction formula
(Chandrasekhar 1943; Binney \& Tremaine 1987), assuming that at the sub-kpc spatial scale, density gradients are not large. We have
considered the secondary BH moving  in the remnant disk plane and decaying slowly enough to be considered at any given moment on a
circular orbit. The change in the orbit of the BH is,  
\begin{equation}
 \frac{dr}{dt}= - \frac{rF_{df}}{M_{\bullet,2} \left[ V_c + r\frac{dV_c}{dr} \right] }.
\end{equation}
where $F_{df}$ is the force experimented by the BH due to dynamical friction and $V_c$ is the circular velocity at radius $r$. Such a
formula gives us a lower limit to the BH's decay time scale due to its interaction with dark matter and stars, which corresponds to the
third stage in the coalescence process.

It is worth noting that most BH pairs ($\gtrsim 90\%$) for all the model parameters explored, do not reach the next stage where the
interaction with a gaseous disk occurs, mostly because the time scale needed for the stellar component to bring together the BHs to a
few tens of parsecs is larger than the time available since the LMM. Also, the binary-stellar medium interaction is not a finished
topic, since there have been some recent results claiming that triaxiality or axisymmetry in stellar distributions is enough to
overcome the final parsec problem in gas-free galaxies (Khan \& Holley-Bockelmann 2013). Whilst other studies claim that, given the
binaries hardening rate dependence on the simulation's number of particles, there is not enough evidence to reach any conclusions yet
(Vasiliev, Antonini \& Merritt 2014). 

A number of recent studies have shown that the presence of a gaseous circumbinary disk is highly efficient to drain angular momentum
out from a binary BH (e.g. Dotti et al. 2012, Hayasaki \etal 2009). Given that it is still unknown whether the interaction with stars
might be able to overcome the final parsec problem, and because galaxy mergers funnel copious amounts of gas towards the remnant galaxy
centre, we include as a fourth and last stage in our models the interaction within a gaseous disk as the mechanism responsible for the
binaries' angular momentum loss. The interaction within gaseous environment have the added value of promising EM signatures as we will
discuss in section 7.

\subsection{Evolution within a circumbinary disk.} 
In the fourth and last stage of our model, the binary exchanges angular momentum with a gaseous circumbinary disk at a rate that can be
approximated by the angular momentum exchange at resonances (Goldreich \& Tremaine, 1979; Hayasaki \etal 2009). The mass transfer from
the CBD adds momentum to the binary and might form accretion disks around each BH. We will follow the analytical model calibrated using
SPH simulations suggested by Hayasaki \etal (2009). It is important to acknowledge that vigorous research work on the angular momentum
exchange between the CBD and the binary BH is currently being undertaken by several groups, and details such as, gas equation of state,
CBD fragmentation, and star formation are still uncertain (Roskar et al 2014, Colpi 2014). It is also worth noting that our model is
flexible enough to extend its implementation to include the results of such studies when the discussion is settled. For now the
efficiency of the angular momentum exchange will be locked in phenomenological parameters in our approach. The CBD initial mass is
given by the CBD to total BHs mass ratio, we explore three values $M_{\rm CBD} / M_{\rm BH}=0.1$, $0.01$ and $0.001$. The $M_{\rm CBD}$
grows along the binary's evolution at a rate given by the parameter $f_{mah}= 0.01$ and $0.001$ (see table \ref{parametros}) that
constrains the cosmic mass transfer. Given that typical time scales at this stage are much shorter than the time between the snapshots
of the simulation, we interpolate the MAHs so the new time steps are at least half the characteristic time scale for the evolution of
the binary given by $t_c \simeq 8\times10^{4} ( M_{\rm BH}^{1/2}/M_{\odot}) [yr]$ (Hayasaki 2009, eq. 26). The evolution of the
semi-major axis can be calculated as (Hayasaki 2009)
\begin{equation}
 \frac{a}{a_0} = \left( \frac{\dot{M}_{c}}{\dot{M}_{crit}} \right)^{-2} 
 \left[ 1- \left( 1- \frac{\dot{M}_{c}}{\dot{M}_{crit}}\right) exp\left( \frac{t}{t_c}
\frac{\dot{M}_{c}}{\dot{M}_{crit}}\right)  \right]^2.
\end{equation}
Here $a_0$ is the initial semi major axis and $\dot{M}_{crit}$ is a critical mass transfer rate defined as:
\begin{equation}
 \dot{M}_{crit} = \frac{1.39}{f_{am}}\frac{q}{(1+q)^2} \frac{M_{\rm BH}}{t_c}.
\end{equation}
The semi-major axis $a$ decays with time when $\dot{M}_{crit}$ is larger than the cosmic accretion rate, $\dot{M}_{c}$, while it
increases when  $\dot{M}_{crit} < \dot{M}_{c}$. The parameter $f_{am}$ determines how much the torque of the mass transfer is converted
to the torque of the binary.

As long as material keeps reaching the CBD, this will provide a considerably large amount of mass for the BHs to accrete. We will
consider accretion triggered only from this stage on, with both BHs accreting mass and following a phenomenological approach where both
BH accrete an amount of mass is given by:
\begin{equation}\label{mpunto}
 \begin{array}{c c c}
  \varDelta M_{\bullet,1} = \frac{1}{1+q} + (f_{mah} M_{mah}) + (f_{d} M_{\rm CBD}) \\ 
   \\
  \varDelta M_{\bullet,2} = \frac{q}{1+q} + (f_{mah} M_{mah}) + (f_{d} M_{\rm CBD}) \\   
 \end{array}
\end{equation}
per time step, where $f_{mah}$ is the fraction of the mass $M_{mah}$ gained by the dark halo MAH in a time step, so the product
$f_{mah} M_{mah}$ is the amount of mass that reaches the galaxy centre in a time step, which we have called \textit{cosmic accretion
rate}; and $f_{d}$ is the fraction of the circumbinary disk mass $M_{\rm CBD}$ accreted by the black holes per time step.  It is clear
that the accretion model oversimplifies the complexity of AGN activity, however it is beyond the scope of our work to accurately
investigate AGN activity. 

We follow the dynamical evolution of the initial galaxy pairs population tracking their separation, for the two BH seeding 
prescriptions (O and F) and the 10 parameter sets shown in table \ref{parametros}, so we have 20 different models. Once the binaries
reach separations of $~ 10^{-3}$pc, they are considered binaries likely to decay and coalesce due to the emission of gravitational
waves or GW candidates. In our model the destiny of BH binaries is set by a competition between cosmic accretion that replenish gas and
angular momentum into the CBD ($f_{mah}$)  and the binary ($f_{d}, f_{am}$), versus the angular momentum drain out of the binary by the
CBD torques ($M_{\rm CBD} / M_{\rm BH}$). The CBD viscosity may have either tidal or stochastic origin (Fiacconi \etal 2013), ours is
an effective model, therefore it does not capture the full detail of the gaseous disk physics.
 \begin{table}
 \centering
  \begin{tabular}{c c c c c}
   \hline
model &  $\text{M}_{\text{CBD}}/\text{M}_{\text{BH}}$ & $f_{am}$ &  $f_{mah}$   & $f_{d}$   \\ \hline
1 & 0.1 & 0.01 & 0.01 & 0.001     \\
2 & 0.1 & 0.01 & 0.01 & 0.0001    \\
3 & 0.1 & 0.01 & 0.001 & 0.001    \\
4 & 0.1 & 0.01 & 0.001 & 0.0001   \\

5 & 0.01 & 0.01  & 0.01 &  0.001    \\
6 & 0.01 & 0.01  & 0.01 &  0.0001   \\
7 & 0.01 & 0.01  & 0.001 &  0.001   \\
8 & 0.01 & 0.01  & 0.001 &  0.0001  \\

9 & 0.1 & 0.01  & 0.01 &  0.1       \\
10& 0.001 & 0.01  & 0.01 &  0.001   \\
 \hline
  \end{tabular}
\caption { \small Parameters used for both BH seeding prescriptions O and F. $M_{\rm CBD}/M_{\rm BH}$ is the initial circumbinary disk
to total BH mass ratio, $f_{am}$ indicates the angular momentum fraction that reaches the BHs, $f_{mah}$ is the MAH mass fraction that
goes to the galaxy centre and $f_{d}$ is the fraction of CBD mass that is accreted by the BHs.}
\label{parametros}
 \end{table}
\section{Binary BH host galaxy properties}
Characterizing the types of galaxies likely to host BH binaries can be useful to determine a low frequency GW source location, to
search for peculiar EM signatures, and to elucidate whether such a peculiar EM signature is truly due to a coalescing binary as 
predicted by various scenarios. Furthermore, if we are able to identify the host galaxy, and therefore the GW source position; 
luminosity distance measurements based on GWs would be greatly improved (Arun \etal 2009). It also reduces the parameter space of
search templates, and correspondingly increases the observed signal-to-noise ratio. Identifying and studying plausible binary BHs
host galaxies by their properties and EM signatures is possible now and will teach us about mergers rate, galaxy assembly histories,
accretion and even might help us narrow uncertainties about the possibility of detecting individual GW sources. Here we present our
predictions for the general properties of galaxies hosting BH binaries prone to decay due to the emission of GWs and with separations
$<10^{-3}$pc.

\subsection{Redshift}
\begin{figure}
 \centering
 \includegraphics[width=0.9\linewidth]{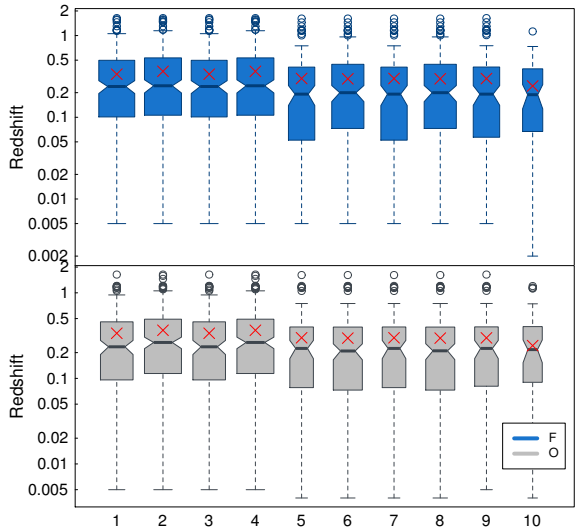}
 \caption{Binaries redshift distributions for all sets of parameters and both BH seeding prescriptions represented by box and whiskers
plots. The box show the interquartile range (IQR) where the central $50\%$ of the data falls. Each whisker or dashed line represents
the lower and upper $25\%$. Darker blue and grey horizontal lines within each box denotes the median values and the notch displays the
$95\%$ confidence interval around the median. Blue and grey open circles represent the outliers outside 1.5 times the interquartile
range. The red crosses represent the mean values. The boxes width is proportional to the square-root of the number of GW candidates. 
We conclude that the redshift distribution is slightly asymmetric with high significance, and half of the systems distributed below
z$\sim$0.21.}
 \label{fig:boxz}
\end{figure}

Regardless of both the model parameters and the BH seeding prescription, the binary BHs redshift distributions, shown in figure
\ref{fig:boxz}, peak around $z \sim 0.21$ with a $95\%$ confidence interval. The distributions reveal an asymmetric shape with 
$\sim75\%$ of the binaries highly concentrated between $0.1 < $ z $< 1$ and a tail towards low redshift including the remaining
$25\%$.  
We note that $\text{M}_{\text{CBD}}/\text{M}_{\text{BH}}$ is the most important parameter determining the redshift distribution.
The distribution peaks at slightly lower redshifts for less massive CBDs (models 5 to 10) due to a less efficient angular momentum
loss mechanism. 

\begin{figure}
 \centering
 \includegraphics[width=0.9\linewidth]{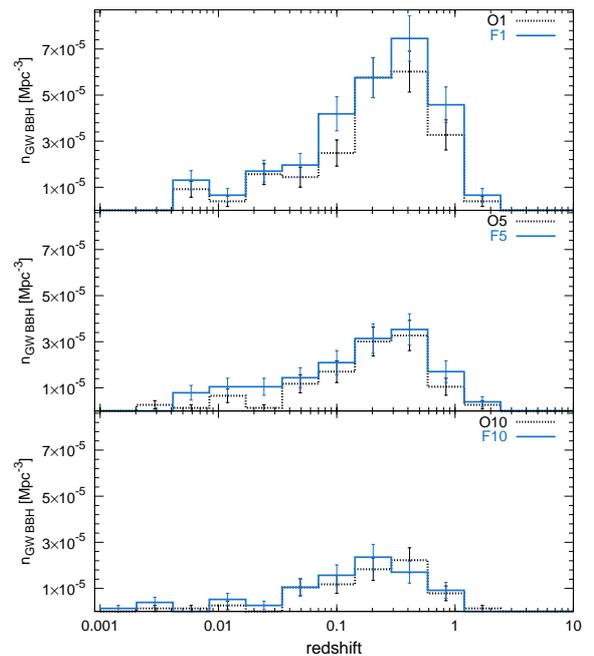}
 \caption{Number of binaries per unit volume likely to reach the GW emission regime as function of redshift. The line colour represents
the BH seeding prescription and each panel corresponds to a parameter set. The errors are Poissonian. Models O1 and F1 with CBDs
initial masses $\text{M}_{\text{CBD}} = 0.1 \, \text{M}_{\text{BH}}$, O5 and F5 have $\text{M}_{\text{CBD}} = 0.01 \,
\text{M}_{\text{BH}}$, and O10 and F10 with $\text{M}_{\text{CBD}} = 0.001 \, \text{M}_{\text{BH}}$. Regardless of the CBD mass
and BH seeding, by observing  $10^6$ galaxies one must find some tens of GW candidates  and an even larger number of kpc scale BH 
pairs.}
 \label{fig:redshift}
\end{figure}

In figure \ref{fig:redshift} we show the number of candidates per unit volume likely to reach the GW emission regime at different
redshifts, for three different values of $M_{\rm CBD}$. Since most of the galaxies had their last major merger at $z\leq1$ and the
large scale characteristics of the binaries are dominated by the cosmology, we see as a consequence that most ($\sim 75\%$) of the
candidates to decay due to GW emission occur at $z\lesssim0.5$. Even though, most pairs ($>90\%$) remain at much larger separations, of
the order of $\sim\text{kpc}$. 

\subsection{Host galaxy halo mass.}
\begin{figure}
 \centering
 \includegraphics[width=0.9\linewidth]{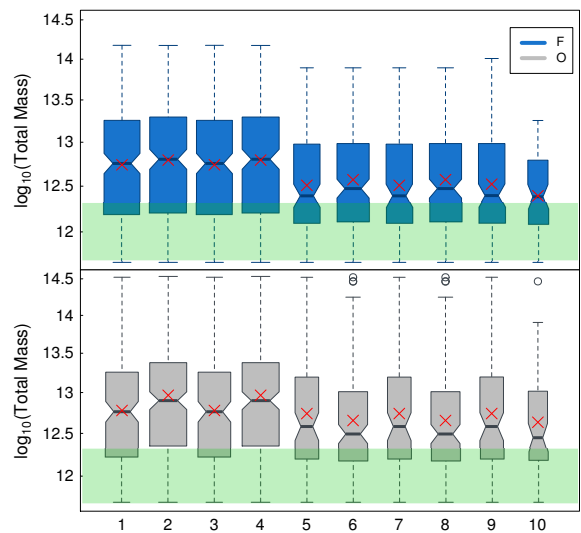}

 \caption{Host galaxy halo mass distributions. Green horizontal region denotes Milky Way halo mass  $1.2^{+0.7}_{-0.4} \text{(stat.)}
^{+0.3}_{-0.3} \text{(sys.)} \times 10^{12} M_\odot$ (68\% confidence) from Busha \etal (2011). Distributions represented by box and
whiskers plots as in figure \ref{fig:boxz}. The distributions are asymmetric with more than $50\%$ of the systems with halo masses
larger than the MW one, with the halo mass mean value depending mostly on the CBD mass, and the high mass tail depending on the BH
seeding scheme.}
 \label{fig:boxmhall}
\end{figure}
 \begin{figure}
  \centering
  \includegraphics[width=0.9\linewidth]{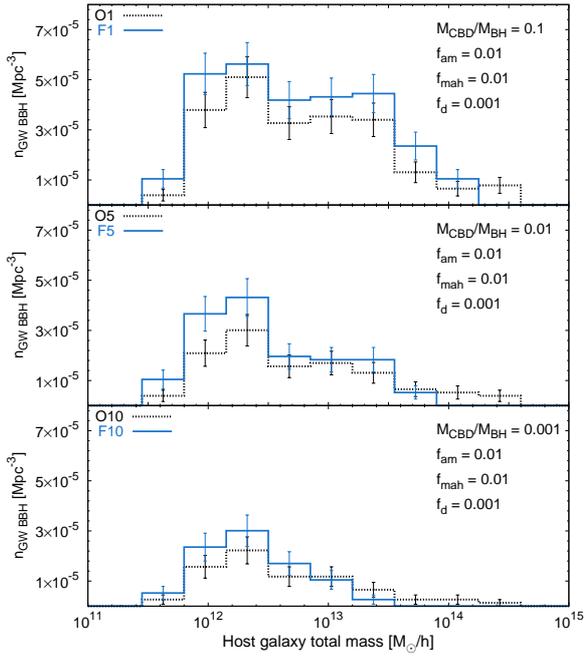}
  \caption{Host galaxy halo mass distribution for three models showing the dependence on $\text{M}_{\text{CBD}} / 
\text{M}_{\text{BH}}$. The errors are Poissonian. It is evident from the figure that $\text{M}_{\text{CBD}}/\text{M}_{\text{BH}}$
is the parameter driving the host galaxy halo mass normalization and shape. However the BH seeding seems to reflect on the high
mass tail.}
  \label{fig:MH}
 \end{figure}
Determining the distribution of dark matter halo masses hosting binary BHs in the GW regime is a first step towards constraining its
origin and abundance. When comparing the host galaxy halo mass distribution for all models, we see from figure \ref{fig:boxmhall} that 
even though there is no agreement on the mass at which the distributions peak across parameter sets, most BH binaries are
systematically found in host galaxies with total mass larger than the Milky Way halo mass. Models with more massive CBD (1-4 in table
\ref{parametros}, with either O and F BH seeding prescription) result in more numerous binary populations, as indicated by the boxes
width (figure \ref{fig:boxmhall}) and the histogram normalization in figure \ref{fig:MH}.

It is clear from figures  \ref{fig:boxmhall} and  \ref{fig:MH} that systems with less massive initial CBD, 
$\text{M}_{\text{CBD}}/\text{M}_{\text{BH}} = 0.01$, tend to peak at lower halo mass, $\sim2.7 \times 10^{12} \text{M}_\odot$ for
F seeding prescription, while for O seeding prescription the more typical host galaxy halo mass for less massive CBD is $\sim3.5
\times 10^{12} \text{M}_\odot$. Unfortunately, total mass estimation is a challenging quantity to measure with current observations,
useful constraints have been set using stacked galaxies plus satellites systems (Yang et al. 2012) or weak lensing (Mandenbaum et al.
2006) but until the next generation of surveys, like LSST, the statistical significance will be limited.

We can see, also from figure \ref{fig:MH}, that $\text{M}_{\text{CBD}}/\text{M}_{\text{BH}}$ is the the driving parameter. If
$\text{M}_{\text{CBD}}$ is smaller the number of BH binaries will be smaller, in particular for those hosted in galaxies with higher
total mass, given the fact their LMM happen at low redshifts (Fakhouri \& Ma 2009) and the binaries are embedded in less massive CBD,
they have less time available to reach the GW emission regime in a non-favourable angular momentum exchange with the CBD. The high mass
tail shows a mild but interesting dependence with the BH seeding scheme, that can be tracked down to the fact that the random nature of
the BH seeding prescription O, may populate high-mass halos with  more massive BHs than the deterministic F seeding scheme. The last
fact assign larger CBD masses and therefore more efficient angular momentum drain from the binary to the O seeding model, pushing some
BH binaries into the GW regime and as a consequence populating the high mass end of the halo mass distributions.

\subsection{Host galaxy stellar mass.}
\begin{figure}
 \centering
 \includegraphics[width=0.9\linewidth]{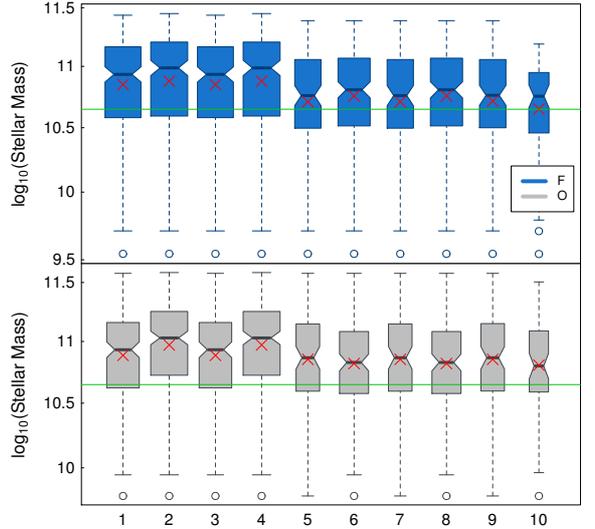}
 \caption{Stellar mass distributions represented by box and whiskers plots as in figure \ref{fig:boxz}. The green line represent  $M^*$
from  Baldry \etal (2008). The average stellar mass of galaxies hosting binary GW candidates is larger than $M^*$ and the distributions
are asymmetric.}
 \label{fig:boxms}
\end{figure}
\begin{figure}
\centering
\includegraphics[width=0.9\linewidth]{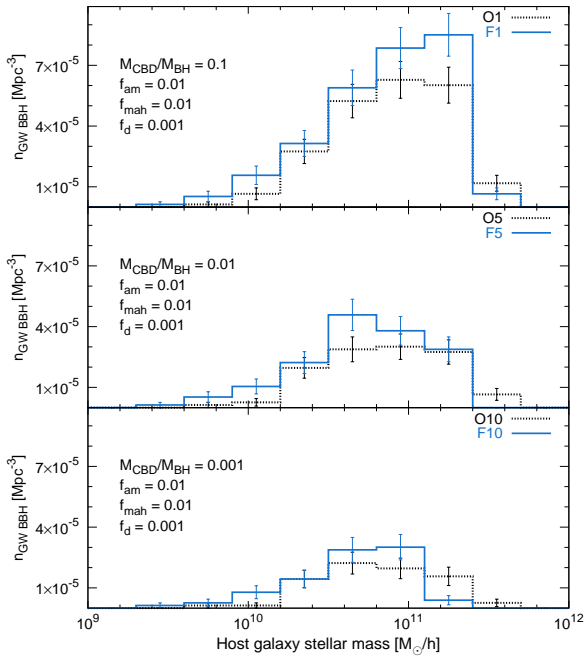}
\caption{Distributions of host galaxy stellar mass for three models showing the dependence on $\text{M}_{\text{CBD}} /
\text{M}_{\text{BH}}$. GW candidates are found in galaxies with stellar mass in the range $10^{10} − 3 \times 10^{11}  M_\odot $. The
errors are Poissonian. The galaxy population hosting GW candidates is more massive in average than $M^*$, and the slope of the host
stellar mass distributions at high values may contain information of BH seeding prescription.}
\label{fig:MS}
\end{figure}
Measurement of galaxies stellar mass, although not free from biases, can be accurately achieved with current galaxy surveys (Baldry
\etal 2008). Thus, it is important and useful to investigate the potential host galaxies stellar mass distribution. Using the halo mass
of the host galaxy, we assign stellar mass following Firmani \& Ávila-Reese (2011) semi-empirical modelling, whose $M_s/M_{\rm H}$
ratio function has a bell shape around $M^*$. A consequence of the dependence on the $M_s/M_{\rm H}$ ratio is that the shape of the
stellar mass distribution is skewed towards low masses. In figures \ref{fig:boxms} and \ref{fig:MS} we can see that  BH binaries host
galaxies have stellar masses distributions with a broad peak in the range $10^{10} - 3\times10^{11} M_\odot$ and a tail in the low mass
end going up to $3\times10^{9} M_\odot$. Most GW candidates ($\sim 75\%$) are hosted in galaxies more massive than $M^* = 4.45 \times
10^{10} \text{M}_\odot$ (Baldry \etal 2008), regardless the model parameters.
  
However for model F10 (fig. \ref{fig:boxms}) the effect of a low mass CBD is more dramatic, binaries in massive host galaxies get stuck
at large  separations, not close enough to enter the GW emission regime. Thus, for this model, the galaxy population hosting GW
candidates is less massive in average with mean stellar mass equal to $M^*$.  The scatter in the BH seeding prescription O, may assign
more massive BHs, and as a consequence more massive CBDs, to the same galaxies than the F prescription, pushing some host galaxies
towards the GW regime. Hence, the slope of the host galaxies stellar mass function at high values may contain information of the BH
seeding.

\section{PTA and SB sources host galaxy properties.}

Recently the possibility of using Pulsar Timing Arrays has been put forward by several groups (Jenet \etal 2009; Manchester \etal 2013;
Ferdman \etal 2010) stimulating studies constraining the binary BH galaxy host abundance in the sky (Simon et al. 2014), suggesting
that the expected abundance of PTA systems is low (Rosado \& Sesana 2013). We split the binary BHs population, and their respective
hosts, in candidates to emit GWs in the PTA frequency range and consider the rest as candidates to emit in the SB gravitational wave
observatories frequency range, such as eLISA, in order to explore consistency with previous PTA studies and to extend the study to SB
systems, which will also increase the number of binary systems prone to show an anomalous EM signature. Typical PTA sources have
$\text{M}_{\text{BH}} \geq 10^8 \text{M}_\odot$ and mass ratios in the range $10^{-3} < q < 1$ (Sesana \etal 2012). 

\subsection{Redshift, halo mass, and stellar mass.}
\begin{figure}
 \centering
 \includegraphics[width=0.8\linewidth]{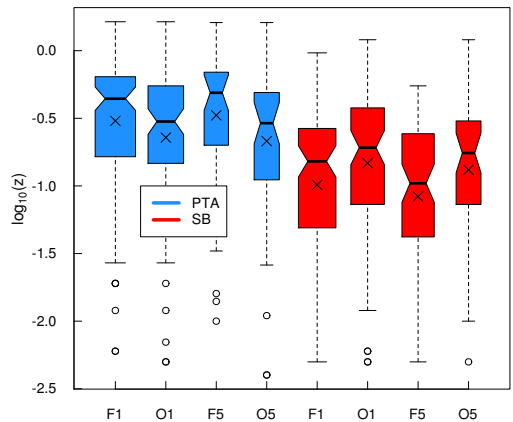}
 \caption{PTA and SB sources redshift distributions in box and whisker plots as in figure \ref{fig:boxz}. As in the general binary BH
host galaxy population both distributions are asymmetric, as a consequence the mean an median are different. An important number of
systems will be low z SB sources.}
 \label{fig:ptaboxZ}
\end{figure}

The redshift distributions for our potential PTA and SB sources are depicted in figure \ref{fig:ptaboxZ}. It shows that most of the
closest PTA sources (given that they have been formed after the last major merger of their host halos/galaxies) peaked at redshift
$0.15<z<0.6$ with a tail going up to $z\sim0.03$. We have found an important population of SB sources, peaking at $0.06<z<0.33$. For
some models, even $50\%$ of the GW candidates might be SB sources that would represent an interesting case for multimessenger
astrophysics in the Local Universe. 

\begin{figure}
 \centering
 \includegraphics[width=0.9\linewidth]{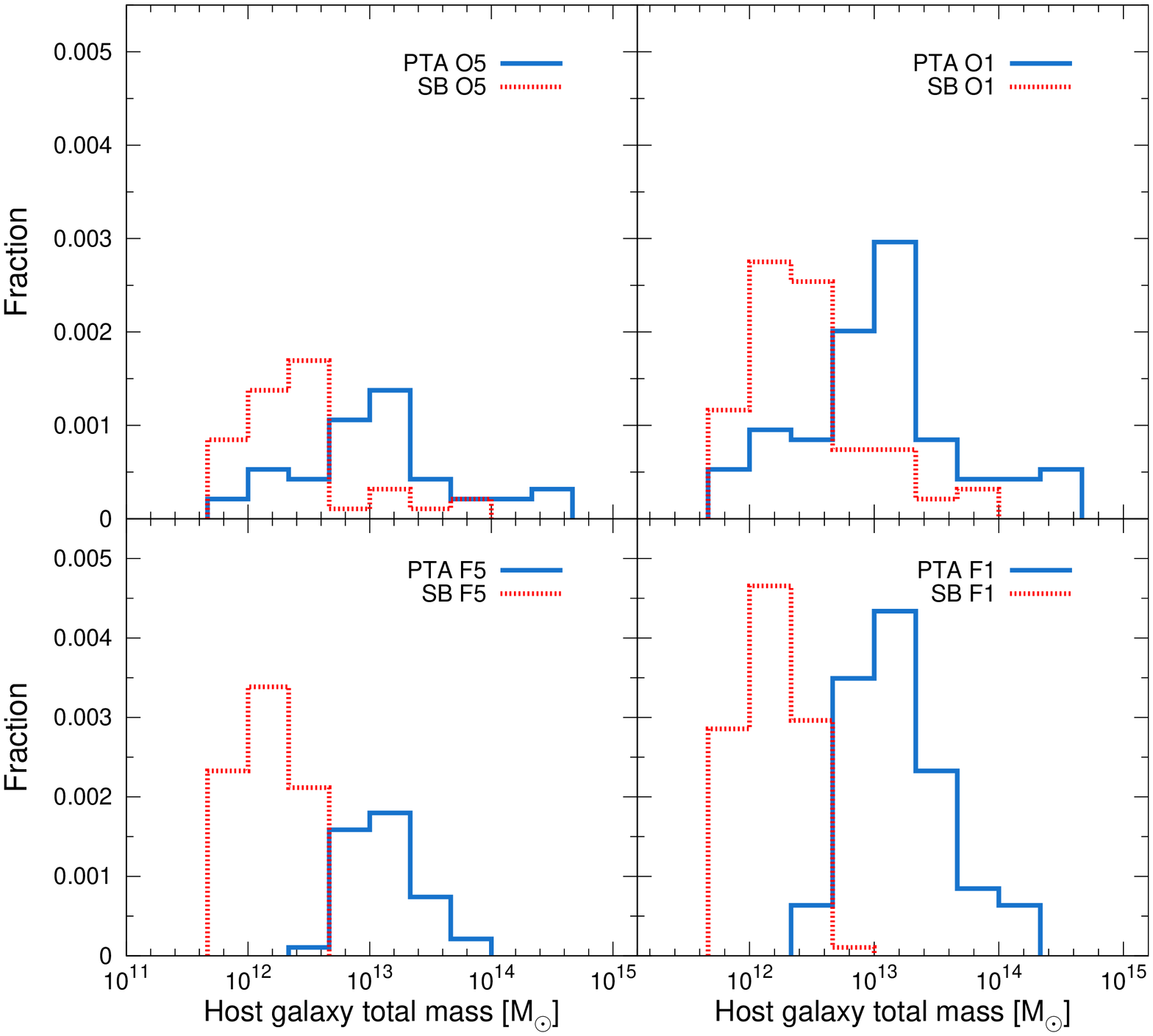}
 \caption{Host galaxy halo mass distribution of PTA detectable GW candidates (blue lines) and space-based (SB) detectable GW candidates
(red lines) for 4 different models. Normalized to the number of galaxies with stellar mass $\geq 1.5 \times 10^{8} M_{\odot}$. Means
and confidence intervals are compare in fig. \ref{fig:ptabsbox}. The expected typical dark halo mass for PTA and SB sources are
segregated in a bimodal distribution.}
 \label{fig:PTAtot}
\end{figure}
\begin{figure}
 \centering
 \includegraphics[width=0.9\linewidth]{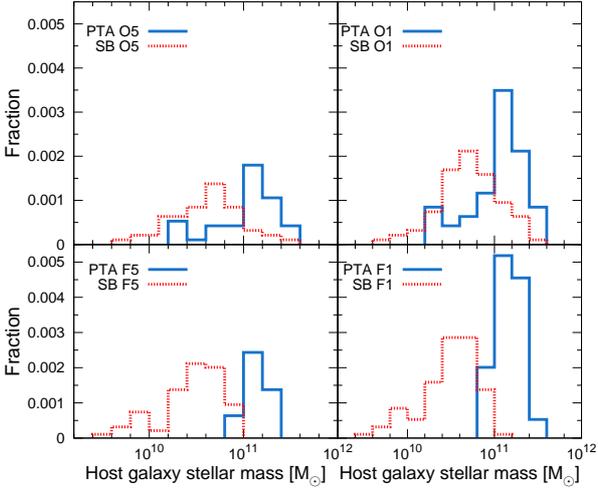}
 \caption{Host galaxy stellar mass distribution of PTA detectable GW candidates (blue lines) and space-based (SB) detectable GW
candidates (red lines) for 4 different models. Normalized to the number of galaxies with stellar mass $\geq 1.5 \times 10^{8}
M_{\odot}$. Means and confidence intervals are compared in fig. \ref{fig:ptabsbox}. The stellar mass distributions also present a
bimodal behaviour.}
 \label{fig:PTAstar}
\end{figure}
\begin{figure}
 \centering
 \includegraphics[width=0.8\linewidth]{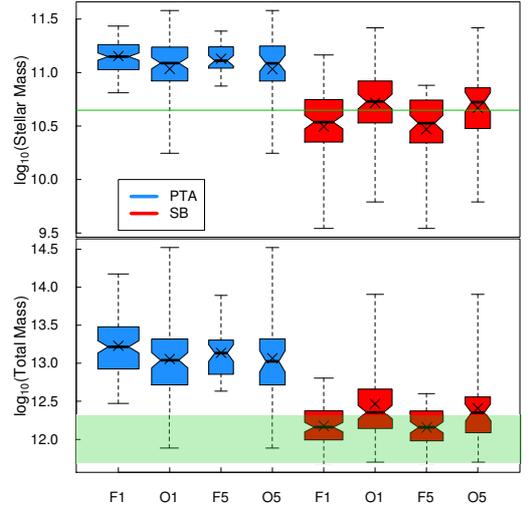}
 \caption{Host galaxy properties for PTA and SB candidates. Stellar mass distributions for 4 models for SB and PTA candidates. The
green line shows $M^*$ (Baldry \etal 2008) (Upper panel). Total mass distributions for galaxies hosting PTA and SB detectable sources.
The green horizontal region denotes Milky Way halo mass (Busha \etal 2011) (Lower panel). In both panels, black lines represent medians
with $95\%$ confidence intervals. Black crosses indicate the mean values. We have estimated that SB sources would be  hosted by
galaxies with $M^*$ in the Local Universe. }
 \label{fig:ptabsbox}
\end{figure}

The host dark halo and stellar masses distributions for SB and PTA binaries are shown in figures \ref{fig:PTAtot}, \ref{fig:PTAstar}
and \ref{fig:ptabsbox}. We observe a bimodal distribution for both quantities. Given the nature of the F seeding prescription (equation
\ref{Fbhmass}) we expect a clear separation between galaxies hosting PTA and SB binaries. However, we observe a similar segregation for
O seeding prescription. The distributions peak around the same mass range for PTA and SB candidates (figure \ref{fig:PTAtot}),  we
conclude that such distributions are due not only to the BH seeding process but mostly to the galaxy pair dynamical evolution. The
bimodal distribution shows the following characteristics:

PTA candidates total mass host galaxy distributions have a broad peak between $5 \times 10^{12} - 3 \times 10^{13} \, \text{M}_\odot$,
with a median in a $95\%$ confidence interval around $10^{13} \text{M}_\odot$ for all models and stellar mass distributions
consistently peaking in the range $9 \times 10^{10} -\, 2\times 10^{11} \text{M}_\odot$ with a median $1.3 \times 10^{11}
\text{M}_\odot$ within a $95\%$ confidence interval as seen in figures \ref{fig:PTAstar} and \ref{fig:ptabsbox}. 

SB detectable sources are found in galaxies with dark halo mass peaking in the range $9.6 \times 10^{11} -\,4.5 \times
10^{12}\,\text{M}_\odot$ with no agreement in the median across parameter sets and average stellar masses between $2\times 10^{10}$ and
$8\times 10^{10} M_\odot$ Given that O seeding prescription allows for lighter BH masses in more massive galaxies which results in a
larger spread in the host halo mass distributions, O models have a slightly higher median when compared to F models. As shown in the
upper panel in figure \ref{fig:ptabsbox} most galaxies containing SB sources (over $50\%$ and up to $75\%$, depending on the parameters
and BH seeding prescription) have average stellar masses between $2 \times10^{10} $ and $8 \times10^{10} \text{M}_\odot$. 

SB candidates are more likely to dwell in Milky Way-like haloes with stellar mass similar to $M^*$. Whilst, galaxies hosting PTA
candidates are in average an order of magnitude more massive than galaxies containing SB candidates. Such distribution bimodality might
be exploited in order to constraint the nature of a set of binary BHs candidates.

\subsection{Environment.}
Host galaxies' environment is yet another key piece of information that may contribute to improve the search strategies of GW
candidates, specifically for the ones with EM anomalous signatures. In order to characterize the host galaxy environment of each binary
candidate to coalesce due to GW emission in the PTA or SB frequency window, we estimate environment overdensities on two different
length scales for the number density of galaxies with masses larger or equal to $10^{10}\,\text{M}_{\odot}h^{-1}$. The overdensity
estimator $\delta_r$ on spheres of radius $r = 5$ and $1.5$ Mpc $h^{-1}$
\begin{equation}
 \delta_r = \frac{\rho_r - \rho_a}{\rho_a}
\end{equation}
is used, where $\rho_a$ is the average number density of galaxies with mass larger or equal to $10^{10} h^{-1} M_{\odot}$ counted from
the whole simulation volume, and $\rho_r$ is the number density in a sphere of radius $r$ centred in each candidate. $\rho_r$ does not
take into account the galaxy where the sphere is centred, i.e. the host galaxy of the binary. A galaxy in an environment with cosmic
mean density has $\delta_r =0$. If $\delta_r= -1$, the galaxy is isolated to the radius r. If $-1 < \delta_r < 0$ the galaxy is within
an underdensity.
\begin{figure*}
 \centering
 \includegraphics[width=0.9\linewidth]{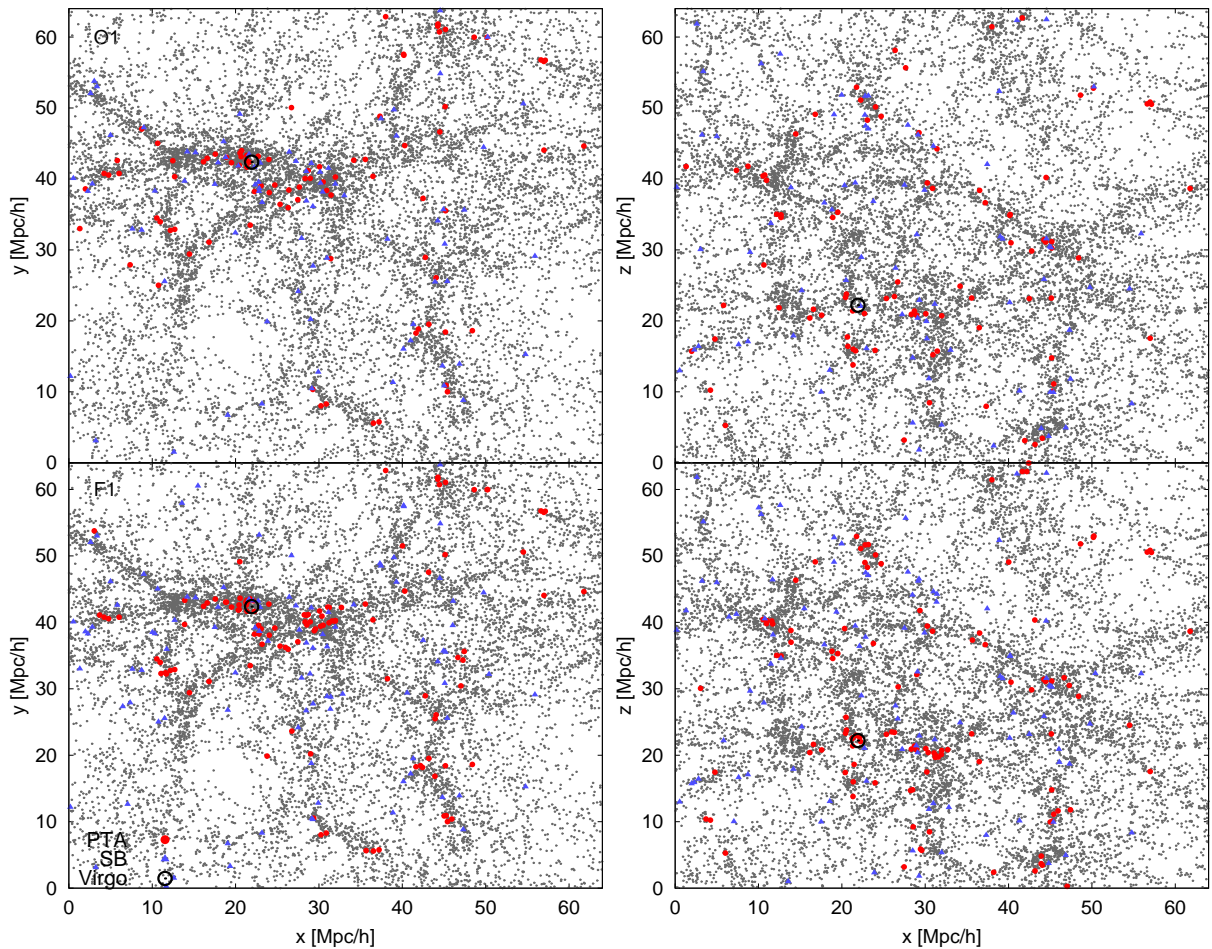}
 \caption{GW candidates projected distribution in the Local Universe simulation for both BH seeding prescriptions. Top panel:  O1 model
candidates. Bottom panel: F1 model candidates. Red points correspond to PTA sources, blue crosses correspond to SB sources. Grey dots
show all galaxies with total masses larger or equal to $2 \times 10^{10} h^{-1}\text{M}_{\odot}$.} 
 \label{fig:PTA-mapa}
\end{figure*}
From the map of PTA and SB projected positions it is possible to investigate the typical environment in which BH coalescences take
place. We conclude that there is a relationship between the environment and the BH seeding prescriptions (figure \ref{fig:PTA-mapa}).
Thus, binary BHs detections combined with environmental information might offer insight on the occupation of black holes and therefore
on BH formation scenarios.
The overdensity for spheres with radii $5$ and $1.5 h^{-1}$Mpc as a function of halo and BH mass, averaged over logarithmic bins with
size $\text{log}_{10}(M_{\rm H}) = \text{log}_{10}(M_{\rm BH}) = 0.5$, is depicted in figure \ref{fig:overtot} and shows that the
overdensity as a function of BH mass shows a clear tendency for more massive BHs, PTA sources in our case, to inhabit more dense
environments. However, the overdensity as function of halo mass (figure \ref{fig:overtot}, bottom panel) shows such a clear separation
only when the BH-halo occupation is given by an increasing power law, as in the case of F models. For O models, whose BH-halo
occupation is given by a broader distribution, we found potential SB and PTA sources in galaxies within similar environments making the
environment an unreliable parameter to determine a binary host when used on its own.

Despite the previous discussion, when inspecting the overdensity distribution for PTA and SB sources as shown in figure 
\ref{fig:overbox} we have found that, regardless the differences introduced by the BH seeding prescriptions, PTA hosts dwell in denser
surroundings with a median $\delta_{n,1.5} = 11.75$ consistent for all models in a $95\%$ confidence interval. SB have overdensity
distributions peaking around $\delta_{n,1.5} = 5.2$. Even though the properties of host galaxies overlap in some cases depending mostly
on the BHs demographics, there are galaxy populations with both, galaxy properties and environments, that might be separated in order
to improve the likelihood of identifying the GW host sources. 
\begin{figure}
 \centering
 \includegraphics[width=0.99\linewidth]{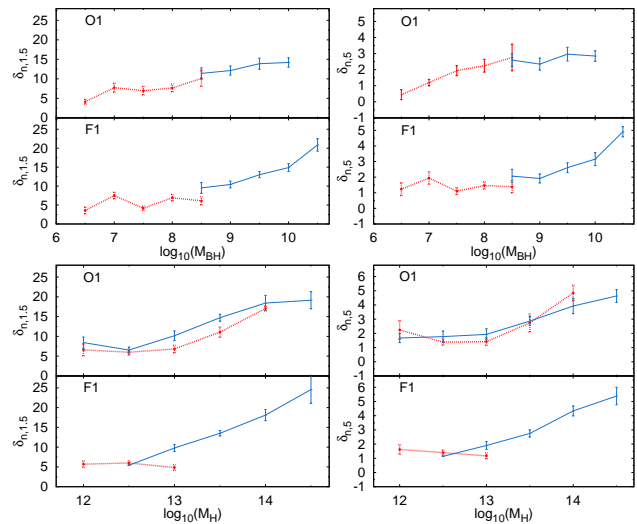}
 \caption{Average overdensity for spheres with radii $5$ and $1.5 h^{-1}$Mpc as a function of halo and BH mass, grouped in
$\text{log}_{10}(M_{\rm H}) = \text{log}_{10}(M_{\rm BH}) = 0.5$ bins. Blue solid lines correspond to PTA sources and red dotted
lines correspond to SB sources. Standard errors for each bin are indicated by the $1\sigma$ error bars.} 
 \label{fig:overtot}
\end{figure}

\begin{figure}
 \centering
 \includegraphics[width=0.9\linewidth]{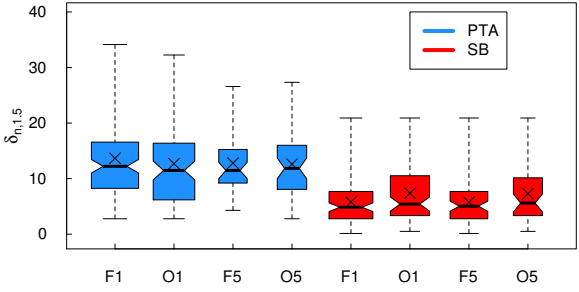}
 \caption{Overdensity of galaxies with mass $\geq 10^{10} \text{M}_\odot h^{-1}$ in a sphere of $1.5h^{-1}$Mpc for PTA and SB sources
and 2 different models. Even though both kinds of binaries might be found in any environment there is a clear tendency for PTAs to be
found in denser environments regardless of the model.}
 \label{fig:overbox}
\end{figure}

\section{Electromagnetic signature. Lifespan and variability}
If the binary hardening happens in a gas rich environment, as we are assuming, circumbinary and individual accretion disks around the
BHs may form. Whilst on the fourth stage of the model, we explore the possible electromagnetic (EM) signatures associated to the CBD
and accretion disks. Following Sesana \etal (2012), by calculating the angular momentum at the inner CBD radius and the size of the
possible accretion disks at each time step for all binaries in our sample we can identify the possible EM emission scenarios. The
change in angular momentum at the CBDs inner radius is given by Hayasaki \etal (2009): 
\begin{equation}
  \dot{J}_{cbd} \simeq 3 \left( \frac{m+1}{l}\right)^{1/3}  \frac{(1+q)^2}{q}  \frac{M_{\rm CBD}}{M_{\rm BH}}
\frac{1}{\sqrt{1-e^2}} \frac{J_b}{\tau_{vis}^{cbd}}.
 \end{equation}
We numerically integrate at every time step in order to get the radii for the Keplerian orbit $r_i = J/GM_{\bullet,i}$ at which the
infalling material is captured around each BH given by the angular momentum of the material in the inner orbit of the CBD. Following
Sesana \etal (2012) and Rees (1988) if $r_i<3R_S$, (with $R_S$ the Schwarzschild radius) the inflow is radial and there will be
Bondi-like accretion. For larger values, the infalling material settles at radius $r_i$ around $M_{\bullet,i}$ and dissipates its
angular momentum through viscous processes, eventually being accreted by the BHs. The time needed for an annulus of material at radius
$r$ to be accreted for an $\alpha$ disk (Shakura \& Sunyaev 1973) is
\begin{equation}
 t_{acc, i} = 0.837\: \text{yr}\: \left( \frac{\dot{M}_i}{\epsilon} \right)^{-2} \frac{ r^{7/2}_{10} M_{8,i}}{\alpha},
\end{equation} 
where $\alpha=0.1$ is the viscosity parameter, the radiative efficiency $\epsilon=0.1$, for all binaries, $r_{10}$ is the accretion
disk size in $10\:R_S$ units of the $i$-BH. If $t_{acc}$ is longer than the binary orbital period $P$, then a lasting, periodically fed
accretion disk will form around the BH. If $t_{acc}$ is shorter than $P$, then accretion is periodic without lasting accretion disks,
which by itself has the distinctive signature of variable accretion associated with the binary period. The radius at which the
transition between this two possibilities happens is found by equating both, the period $P$ and the accretion time $t_{acc}$.
\begin{equation} \label{eq:REM}
 r_{\text{trans}} = 5.3 \: R_S\: \alpha^{2/7} \left(  \frac{\dot{M}_{i}}{\epsilon}
\right)^{4/7} M^{-2/7}_{8,i} P^{2/7}
\end{equation}
where $P$ is the binary period in years, $\dot{M}_{0.3,i}$ is the accretion mass rate as defined by equation \ref{mpunto} and
$M_{8,i}$ is the mass of the $i$-BH in units of $10^8 M_{\odot}$. 

It is possible to identify 3 scenarios for EM signatures for binaries that are still coupled to the CBD and a spectral feature in
binaries that have decoupled. We explore the abundance of the following EM signatures in our complete binary population and  its
dependence on the BH seeding prescription.
\begin{enumerate}
 \item  If $r_i < r_{\text{trans}}$ or equivalently $t_{acc,i} < P$, there will be steady optical/IR emission from the
CBD. Material falling from the CBD inner radius will form an accretion disk around the BH, whose accretion time will be
short compared to the orbital period, thus presenting strong periodicity in the UV/soft X-ray. 

 \item If $r_i > r_{\text{trans}}$ or $t_{acc,i} > P$ there will be steady UV/soft X-ray emission due to the long
lived accretion disk around  the BH, which is likely to produce a relativistic Fe $\text{K}_{\alpha}$ line. Thus, in
case this condition is met by the two BHs, the two Fe $\text{K}_{\alpha}$ superposed lines might be observable, in
addition to the optical/IR emission from the CBD.

 \item If $r_i < 3R_S$, there is radial accretion. No signatures of an accreting system are present, only the  optical/IR 
emission from the CBD.  

 \item If $P < 1 $ yr and the binary is decoupled from the CBD, the combined spectrum from the CBD plus accretion disk
might show a deficient soft X-ray continuum and UV emission compared with a single BH of mass $\sim M_{\rm BH}$. Such
deficient emission being more evident for shorter periods.
\end{enumerate}

Although the aforementioned features have been proposed by Sesana \etal (i,ii,iii; 2012) and Tanaka \etal (iv, 2012) for PTA sources,
we have extended their application to less massive binaries. We keep track of the time during which the conditions for one of the EM
signatures proposed are met and consider this the lifespan of the associated signature. As the binary's angular momentum and separation
decreases due to the interaction within gaseous disks, the GW driven coalescence takes over, such mechanism is not included in the
model current version. Therefore, the electromagnetic signature lifespan found here might be considered as an upper limit, since the
orbital decay is much quicker once the GW emission drives the angular momentum loss (Peters, 1964; Heiman \etal 2012).

Almost all SB sources have a 3-disk configuration whilst on the last stage, i.e. EM signature type ii: A CBD and two long lived
accretion disks around each BH with viscous time scale longer than the period, resulting in a EM signature given by an optical/IR
continuum from the CBD and Double X-ray Fe $K_\alpha$ emission lines from the accretion disks. While accretion variability might not be
associated with the period, shocks between the infalling material and the accretion disks may produce X-ray hot spots varying on the
orbital period, most SB sources have periods of $1.25 - 18$ yr. 

\begin{figure}
 \centering
 \includegraphics[width=0.9\linewidth]{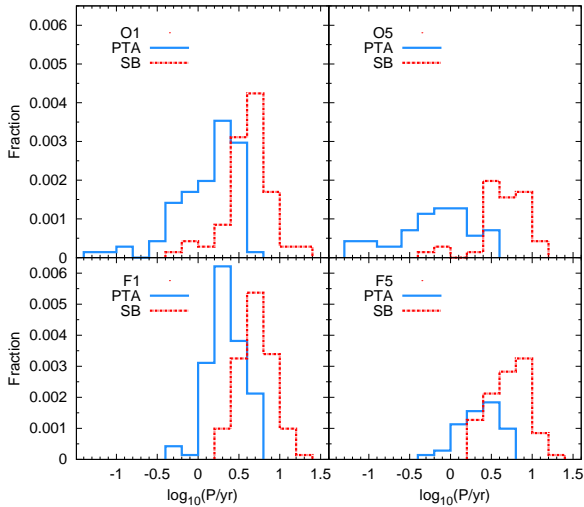}
 \caption{Period of the binaries when they enter the GW decay stage. The blue distribution describes the binaries that emit GW in the
PTA detection frequency window. The red distribution correspond to the binaries emitting in the space based detector frequency window.
Normalized to the number of galaxies with stellar mass $\geq 1.5\times10^8 \text{M}_\odot$}
 \label{fig:period}
\end{figure}

In figure \ref{fig:period} we see that SB sources in haloes populated with F seeding prescription (empirical with increasing power law
behaviour) do not show binaries with $P < 1$ yr. For binaries with O seeding prescription there are a few binaries with period equal or
shorter than a year, the more massive ones allowed by the broader dispersion in the seeding prescription. Thus, less than $\sim8\%$
binaries with different parameter values (O seeding prescription only) might show the EM signature type iv, abnormally weak soft X-ray
continuum and UV emission as proposed by Tanaka \etal (2012). 

\begin{figure}
 \centering
 \includegraphics[width=0.9\linewidth]{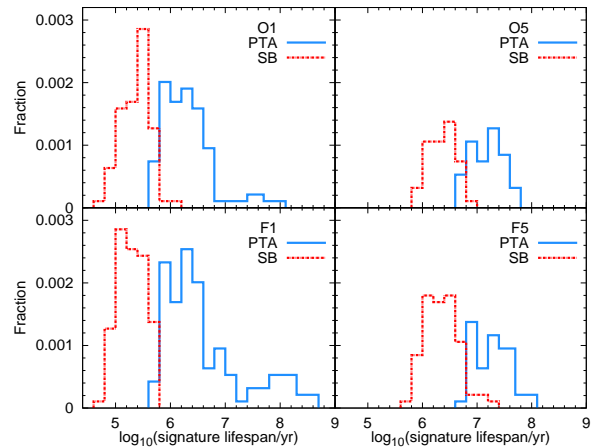}
 \caption{Fraction of SB and PTA sources per number of galaxies with stellar mass larger or equal to $1.5\times 10 M_\odot$ as a
function of the time spent within a CBD. This time scale might be considered an upper limit of the EM signature lifespan.}
 \label{fig:lifespan}
\end{figure}

We found that the lifespan of the EM signature due to a 3-disk configuration ranges between $10^5 - 10^9$ years (figure
\ref{fig:lifespan}) compatible with the theorized AGN duty-cycle. For less massive CBDs (models 5,6,7,8 \& 10), the angular momentum
loss decreases, producing less pairs reaching the GW regime. Those that are able to reach such regime will spend more time interacting
with the CBD resulting in a longer lifespan for the EM signature. 

As in the case of SB sources, most PTA sources, regardless the model, show the dynamical conditions to have a 3-disk configuration with
a EM signature type ii, with some differences. Models F10 and O10 (our models with the less massive CBD) are dominated by signature
type i, same configuration suggested by Roedig \etal (2011) simulations, which consists of two accretion disks with viscous time scale
shorter than the period whose signature might be observed as accretion associated variability in the UV/Soft X-ray.

For models $5-8$ (for both seeding prescriptions, O and F) $\sim 3-4\%$ of the binaries do not fulfill the conditions to form accretion
disks, so a small fraction of binaries would not be detected as an accreting system.  

According with our model, PTA sources have periods within the range $0.4 \sim 6$ yr, figure \ref{fig:period}, with a tail going up to a
few days for some models. Indicating that those binaries with a long lived accretion disks might show variability associated to shocks
caused by streaming material into the accretion disk coming from the CBD or if the  typical CBD in nature turn out to be light (as in
the case of models O10 and F10) there might be an accretion associated emission with variability of a few years.

\section{Summary and conclusions}

We have developed a dynamical model for the decay of supermassive black hole binaries. It has been applied to create a population of
binaries likely to decay due to GW emission within a cosmological volume similar to the Local Universe. 

Our model is coupled with a cosmological simulation, we retrieved the redshift of the last major merger for all haloes with masses
$M_{\rm H} \geq 1.64 \times 10^{10} M_\odot h^{-1}$ (at $z=0$), this zLMM constitutes the starting point for the binary evolution.
We used semi-empirical approaches to assign galaxies and BHs to the haloes. 
In the final stage, the interaction within a gaseous disk is responsible for the binaries' coalescence. The binary's destiny is set by
a competition between cosmic accretion that replenish gas and angular momentum into the CBD and the binary, versus the angular momentum
drain out of the binary by the CBD torques. 

For the binaries found we used a simple criteria based on BHs mass and separation to create a population of binaries prone to emit GW
in the PTA frequency window and in that of the space-based detectors. We found a bimodal correlation between host galaxy properties,
environment, and binary period. Such correlation defines the properties that the host galaxy populations of PTA or SB sources might
show. We anticipate that individual cases are rather difficult to distinguish between binaries and anomalous AGNs, but at a demographic
level it might be possible to identify a subset of AGNs powered by binaries in the subpc regime.  

We investigate the BH binaries host galaxies properties, environment characterization and likelihood to observe some of the EM
signatures previously proposed. A short summary of our results follows here:

\begin{itemize}
 \item In our model for the binaries in the gaseous disk stage the most dominant parameter is the $ M_{\rm CBD}/M_{\rm BH}$ ratio. The
CBD mass works as a regulator of the angular momentum exchange mechanism for the whole binary populations across models, overpowering
even the cosmic accretion rate.

 \item The expected number density of binaries (SB and PTA sources combined) in the Local Universe is $10^{-4} h^3\text{Mpc}^{-3}$.
With a probability of finding a binary every 1000 galaxies with stellar masses larger or equal to $1.5\times10^{8} \text{M}_{\odot}$ in
our best case scenario.

 \item Most of the candidates to decay due to GW emission occur at redshifts between $0.1 -\, 0.5$, as seen in figure
\ref{fig:redshift}, with $\sim 80\%$ occurring at $z\lesssim0.5$ (fig. \ref{fig:boxz}). The distribution peaks at lower redshifts
for less massive CBDs due to the lack of an efficient angular momentum loss mechanism, but the tendency towards low redshifts remains
regardless of model parameters and seeding prescription.

 \item GW candidates host galaxies have typical stellar masses between $10^{10}$ - $3\times10^{11}$ $\text{M}_{\odot}$
(fig.\ref{fig:MS}), with $\sim75\%$ of the candidates having masses larger or equal to $\text{M}^* = 4.45 \times 10^{10}
\text{M}_{\odot}$.

  \item Most GW candidates are systematically found in host galaxies with total masses larger than the Milky Way halo mass
$1.2^{+0.7}_{-0.4} \text{(stat.)} ^{+0.3}_{-0.3} \text{(sys.)} \times 10^{12} M_\odot$ (Busha \etal 2011).
  
 \item As expected, PTA detectable GW candidates occupy the most massive galaxies (in agreement with Rosado \& Sesana, 2013; Simon
\etal 2014) with mean total mass $\sim 10^{13} \text{M}_{\odot}$ and mean stellar mass $\sim 1.4\times  10^{11} \text{M}_{\odot}$. They
are located on average in overdensity regions of $\delta_{n,1.5} = 11.75$, about 10 times the average density of the Universe.

\item Most PTA sources might show EM signatures of type ii, long lived CBD plus accretions disks, with emission associated to shocks
between the accretion disks and streaming material from the CBD with variabilities in the range $1.25 - 20$ yr peaking at $\sim3$ yr
with a lifespan for the EM signature due to the 3 disk configuration between $10^6 - 10^9$ yrs. There is a good chance ($\sim 50 \%$)
in particular for O seeding prescription, to show a deficient soft X-ray continuum and UV emission compared to 'normal' AGNs powered by
one BH as suggested by Tanaka \etal (2012).

 \item According to our model, SB sources might be found in galaxies with total mass peaking in the range $9.6\times10^{11} - \,
4.5\times10^{12} M_\odot $, with average stellar masses between $2 \times10^{10}$ and $8\times10^{10} M_\odot$. Dwelling in
overdensities of $\delta_{n,1.5} = 5.2$,  less dense than the typical environment of PTA sources. 

\item Most SB sources might show EM signatures associated to shocks between the accretion disks and streaming material from the CBD
with variabilities in the range $1.25 - 20$ yr peaking at $\sim3$ yr with a lifespan for the EM signature due to the 3-disk
configuration (EM signature type ii) between $10^5 - 10^6$ yrs and $10^6 - 10^7$ yrs for less massive CBDs.

\end{itemize}

The correlations found may serve as a strategy to distinguish binary local effects in AGNs using a combination of host galaxy
properties and EM signals. The distributions of the binary properties depends on the BHs seeding prescription, thus a sample of
binaries would represent an independent way to learn about BHs formation models and initial BHs population, a possibility that should
be explored. A better understanding on the mechanisms driving the evolution and final coalescence of BH binaries might be reached when
better observational constraints on the number density of BH binaries are available. This will help constrain the parameters used in
this and similar models.

\section{Acknowledgements} 
We thank the CLUES collaboration (www.clues-project.org) for providing access to the simulation analysed in this paper. Our
semi-analytical modelling and analyses were done using IA-UNAM cluster Atocatl.  EMP thanks J. A. de Diego, V. Ávila-Reese, W. H. Lee,
Y. Krongold and A. Batta for helpful conversations. This work was supported by UNAM-PAPIIT grant IN112313 and a graduate CONACyT
fellowship.

%
%
%
%
%

\label{lastpage}

\end{document}